\documentclass[preprint, prb]{revtex4}%
\usepackage{amsfonts}
\usepackage{amsmath}
\usepackage{amssymb}
\usepackage{graphicx}%
\providecommand{\U}[1]{\protect\rule{.1in}{.1in}}
\begin{document}
\title[ ]{Interference Between Source-Free Radiation and Radiation from Sources:
Particle-Like Behavior for Classical Radiation}
\author{Timothy H. Boyer}
\affiliation{Department of Physics, City College of the City University of New York, New
York, New York 10031}

\begin{abstract}
A simple junior-level electrodynamics problem is used to illustrate the
interference between a source-free standing plane wave and a wave generated by
a pulse in a current sheet. \ Depending upon the relative phases between the
standing wave and the current pulse and also upon the relative magnitudes, we
can find quite different patterns of emitted energy and momentum. \ If the
source gives a large radiation pulse so that the source-free plane wave can be
neglected, then the radiation spreads out symmetrically on either side of the
current sheet. \ However, if the radiation sheet gives a pulse with fields
comparable to those of the standing wave, then we can find a single radiation
pulse moving to the right while the current sheet recoils to the left, or the
situation with the directions reversed. \ The example is a crude illustration
of particle-like behavior arising from conventional classical electromagnetic
behavior in the presence of source-free radiation. \ The discussion makes
contact with the ideas of photons in modern physics.

\end{abstract}
\maketitle

\section{Introduction}

Although textbooks of classical electrodynamics sometimes discuss the
scattering and interference of electromagnetic waves in the context of optical
behavior, there seems to be little attention paid to the interference arising
between a source-free wave and the wave emitted by a transient current source.
\ This omission is parallel to the failure of today's textbooks of classical
electromagnetism to include the source-free solution of Maxwell's equations in
addition to the radiation due to a current source when presenting the general
solution of Maxwell's equations. \ Inclusion of the source-free solution opens
the possibility of understanding further aspects of electromagnetic
technology, and also of understanding aspects of modern physics which are not
usually treated in terms of conventional classical electromagnetic theory.
\ Here we give illustrations of interference between source-free radiation and
radiation from a current source which is appropriate for a junior-level course
in electromagnetism. Such interference can give rise to behavior which has
some of the aspects usually associated with wave-particle duality and may give
insight into photon-like physics. \ \ 

We begin with a discussion of the general solution of Maxwell's equation which
contains both a source-free term and a source term involving charges and
currents at a retarded time. \ Next we treat an oscillating current sheet in
the absence of source-free radiation. \ We give the plane wave emitted by the
oscillating surface current and confirm the balance in power delivered by the
source and the outflow of energy in the waves. \ Next, we consider a pulsed
current sheet, and again confirm the energy and momentum conservation ideas
for the pulses moving symmetrically outwards from the surface current. \ At
this point, we introduce a source-free solution of Maxwell's equations
corresponding to a standing plane wave throughout space at the same frequency
as the frequency of the surface current but with arbitrary amplitude and
arbitrary phases in both space and time. \ We show that the energy and
momentum delivered by the pulsed current sheet now depend upon the presence of
the source-free standing wave. \ By varying the phases and amplitude of the
source-free standing wave relative to the wave emitted by the pulsed source,
we find a variety of behavior, some of which suggests particle-like aspects.
In the final section, we remark on the connections of these simple
calculations with the ideas of modern physics.

\section{Source-Free Radiation and the General Solution of Maxwell's
Equations}

The general retarded solution of Maxwell's equations can be written in terms
of the scalar and vector potentials as\cite{gaussian}%
\begin{equation}
V(\mathbf{r,}t)=V^{in}(\mathbf{r},t)+\int d^{3}r^{\prime}\frac{\rho
(\mathbf{r},t-|\mathbf{r-r}^{\prime}|/c)}{|\mathbf{r-r}^{\prime}|}\label{V}%
\end{equation}%
\begin{equation}
\mathbf{A}(\mathbf{r,}t)=\mathbf{A}^{in}(\mathbf{r},t)+\int d^{3}r^{\prime
}\frac{\mathbf{J}(\mathbf{r},t-|\mathbf{r-r}^{\prime}|/c)}{c|\mathbf{r-r}%
^{\prime}|}\label{A}%
\end{equation}
where $V^{in}(\mathbf{r},t)$ and $\mathbf{A}^{in}(\mathbf{r},t)$ correspond to
\textit{source-free} solutions of Maxwell's equations. \ The authors of the
electromagnetism texts agree that this in the correct form for the expression
for the potentials, but they usually exclude the source-free terms in the
final general-solution equations of their texts.\cite{Griffiths}$^{,}%
$\cite{Jackson}$^{,}$\cite{Zangwill}$^{,}$\cite{Garg} \ The source-free
potentials $V^{in}(\mathbf{r},t)$ and $\mathbf{A}^{in}(\mathbf{r},t)$ are
connected to source-free electric and magnetic fields $\mathbf{E}^{in}=-\nabla
V^{in}-(1/c)\partial\mathbf{A}^{in}\mathbf{/\partial t}$ and $\mathbf{B}%
^{in}=\nabla\times\mathbf{A}^{in}.$ \ 

The universe in which we live is not that of the classical electromagnetism
texts which take $V^{in}(\mathbf{r},t)=0$ and $\mathbf{A}^{in}(\mathbf{r}%
,t)=0.$ \ We live in a universe which is filled with radiation which we do not
control with our local equipment. \ Thus in our laboratory rooms, there are
radio waves from various transmitters (which our students pick-up on their
smart phones), and there is thermal radiation from the walls and objects in
the rooms. \ There is also the random radiation present even at the absolute
zero of temperature which gives rise to Casimir forces between uncharged
macroscopic objects. \ This radiation which we do \textit{not} control with
our local equipment can be regarded locally as a source-free solution of
Maxwell's equation which is present when we turn on our local sources. \ It is
the interference between the source-free radiation and the radiation generated
by the local sources which we discuss in the following examples.

\section{Emitted Radiation from an Oscillating Current Sheet Source and No
Source-Free Radiation}

We start with the familiar situation where the source-free fields
$\mathbf{E}^{in}$ and $\mathbf{B}^{in}$ are taken to vanish. \ A very simple
example of a radiation source is an infinite (charge-neutral) oscillating
current sheet which gives rise to plane waves.\cite{G499} \ In empty space,
the plane waves spread out symmetrically on either side from the current
sheet. \ The power per unit area delivered to the surface current by an
external agent appears in the Poynting vectors of the plane wave pointing away
from the current sheet. \ 

Here we take the surface current as%
\begin{equation}
\mathbf{K(}t\mathbf{)}=\widehat{x}K_{0}\cos(\omega t) \label{K}%
\end{equation}
in the plane $z=0,$ giving rise to two plane waves moving outwards from the
sheet,\cite{gaussian}
\begin{equation}
\mathbf{E}_{r}(\mathbf{r},t)=\widehat{x}E_{0}\cos(kz-\omega t),\text{
\ \ }\mathbf{B}_{r}\mathbf{(r,}t)=\widehat{y}E_{0}\cos(kz-\omega t)\text{
\ for }z>0, \label{Er}%
\end{equation}%
\begin{equation}
\mathbf{E}_{l}(\mathbf{r},t)=\widehat{x}E_{0}\cos(kz+\omega t),\text{
\ \ }\mathbf{B}_{l}\mathbf{(r,}t)=-\widehat{y}E_{0}\cos(kz+\omega t)\text{
\ for }z<0, \label{El}%
\end{equation}
where $\omega=ck,$ and the subscripts $r$ and $l$ correspond to
\textquotedblleft right\textquotedblright\ and \textquotedblleft
left.\textquotedblright\ \ Plane waves are solutions of the source-free
Maxwell equations,\cite{G397} so that we need only check the boundary
conditions at $z=0$ to confirm that we have a valid solution of the full
Maxwell equations. \ At the surface $z=0,$ the electric field $\mathbf{E}$ is
continuous at all times $t,$ and the magnetic field $\mathbf{B}$ in Eqs.
(\ref{Er}) and (\ref{El}) is discontinuous, giving%
\begin{equation}
\frac{4\pi}{c}\mathbf{K(}t)=\widehat{z}\times\lbrack\mathbf{B}_{r}%
(0,t)-\mathbf{B}_{l}(0,t)]=-\widehat{x}2E_{0}\cos(\omega t), \label{bc}%
\end{equation}
so that from Eq. (\ref{K}), we have the relation between the amplitude of the
emitted wave and the amplitude of the surface current
\begin{equation}
E_{0}=-\frac{2\pi}{c}K_{0}. \label{EK}%
\end{equation}

The power per unit area $\mathcal{P}/\mathcal{A}$ delivered by the external
source which is driving the surface current is given by the negative of the
electromagnetic power per unit area delivered by the electromagnetic fields.
\ Then from Eqs. (\ref{K})-(\ref{El}) and (\ref{EK}), we have%
\begin{equation}
\mathcal{P}/\mathcal{A}=-\mathbf{K}(t)\cdot\left[  \mathbf{E}_{r}%
(0,t)+\mathbf{E}_{l}(0,t)\right]  /2=-K_{0}E_{0}\cos^{2}(\omega t)=\frac
{c}{2\pi}E_{0}^{2}\cos^{2}(\omega t),\label{PA}%
\end{equation}
where we have taken the electric field \textquotedblleft at\textquotedblright%
\ the current sheet as the average $[\mathbf{E}_{r}+\mathbf{E}_{l}]/2$ of the
electric fields on the two sides of the sheet. \ The Poynting vector on the
right-hand side of the current sheet is
\begin{equation}
\mathbf{S}_{r}\mathbf{(}0,t)=\frac{c}{4\pi}\mathbf{E}_{r}(0,t)\times
\mathbf{B}_{r}(0,t)=\widehat{z}\frac{c}{4\pi}E_{0}^{2}\cos^{2}(\omega
t),\label{Sr}%
\end{equation}
and on the left-hand side is
\begin{equation}
\mathbf{S}_{l}\mathbf{(}0,t)=\frac{c}{4\pi}\mathbf{E}_{l}(0,t)\times
\mathbf{B}_{l}(0,t)=-\widehat{z}\frac{c}{4\pi}E_{0}^{2}\cos^{2}(\omega
t),\label{Sl}%
\end{equation}
so that the outgoing power per unit area in the waves balances the power per
unit area delivered by the driving source%
\begin{equation}
\mathcal{P}/\mathcal{A=}\frac{c}{2\pi}E_{0}^{2}\cos^{2}(\omega t)=\widehat{z}%
\cdot\mathbf{S}_{r}\mathbf{(}0,t)+(-\widehat{z})\cdot\mathbf{S}_{l}%
\mathbf{(}0,t).\label{PAS}%
\end{equation}

\section{Oscillating Current Sheet and In-Coming Plane Waves}

In the situation above, we considered the current sheet as generating the
out-going plane waves on either side. \ However, if we simply add an
appropriate source-free plane standing-wave solution, then we can obtain the
situation corresponding to in-coming waves which are absorbed by the
oscillating current sheet. \ Thus we introduce the source-free standing wave%
\begin{equation}
\mathbf{E}_{S}(z,t)=-\widehat{x}2E_{0}\cos(kz)\cos(\omega t),\text{
\ \ }\mathbf{B}_{S}(z\mathbf{,}t)=-\widehat{y}2E_{0}\sin(kz)\sin(\omega t)
\label{ES}%
\end{equation}
throughout the spacetime in addition to the retarded fields generated by the
oscillating current sheet. \ Then the full electromagnetic fields are
$\mathbf{E=E}_{S}+\mathbf{E}_{c},$ $\mathbf{B=B}_{S}+\mathbf{B}_{c},$ where
$\mathbf{E}_{c}$ and $\mathbf{B}_{c}$ are the retarded fields generated by the
current sheet, given in Eqs. (\ref{Er}) and (\ref{El}). \ But the standing
wave fields in Eq. (\ref{ES}) can be rewritten as running waves using%
\begin{align}
\mathbf{E}_{S}(z,t)  &  =-\widehat{x}E_{0}\cos(kz-\omega t)-\widehat{x}%
E_{0}\cos(kz+\omega t)\nonumber\\
\mathbf{B}_{S}(z,t)  &  =-\widehat{y}E_{0}\cos(kz-\omega t)+\widehat{y}%
E_{0}\cos(kz+\omega t). \label{cs}%
\end{align}
We see that the out-going plane waves generated by the current sheet as given
in Eqs. (\ref{Er}) and (\ref{El}) serve to cancel the outgoing part of the
standing plane wave leaving only the in-coming part of the standing wave.
\ Thus by adding an appropriate source-free solution to the retarded field
generated by the current sheet, we have obtained an entirely different
situation where the signs of the energy in Eqs. (\ref{PA}), (\ref{Sr}), and
(\ref{Sl}) are all reversed. \ The surface current is now absorbing the energy
of symmetrical in-coming plane waves.

\section{Pulsed Oscillating Current Sheet and No Source-Free Radiation}

We now go back to the situation where the source-free solution of Maxwell's
equations is assumed to vanish. \ If the driving source acts only during a
time interval corresponding to half a cycle so that equation (\ref{K}) holds
not for all times $t$ but rather only for time $-\pi/(2\omega)<t<\pi
/(2\omega),$ then there are two plane-wave \textit{pulses} of radiation
emitted,\cite{G499} so that equations (\ref{Er}) and (\ref{El}) become%
\begin{equation}
\mathbf{E}_{r}(\mathbf{r},t)=\widehat{x}E_{0}\cos(kz-\omega t),\text{
\ \ }\mathbf{B}_{r}\mathbf{(r,}t)=\widehat{y}E_{0}\cos(kz-\omega t)\text{
\ for }-\pi/2<kz-\omega t<\pi/2, \label{Erp}%
\end{equation}%
\begin{equation}
\mathbf{E}_{l}(\mathbf{r},t)=\widehat{x}E_{0}\cos(kz+\omega t),\text{
\ \ }\mathbf{B}_{l}\mathbf{(r,}t)=-\widehat{y}E_{0}\cos(kz+\omega t)\text{
\ for }-\pi/2<kz+\omega t<\pi/2. \label{Elp}%
\end{equation}
We still have the energy balance involving the energy per unit area delivered
by the external agent and the energy per unit area appearing in the external
pulses. \ Thus the energy per unit area $U/\mathcal{A}$ delivered by the
external agent corresponds to the time-integral of the power per unit area in
Eq. (\ref{PA})
\begin{equation}
\frac{U}{\mathcal{A}}=\int_{t=-\pi/2\omega}^{t=\pi/2\omega}dt\frac
{\mathcal{P}}{\mathcal{A}}=\int_{t=-\pi/2\omega}^{t=\pi/2\omega}dt\frac
{c}{2\pi}E_{0}^{2}\cos^{2}(\omega t)=\frac{1}{4k}E_{0}^{2}. \label{UA}%
\end{equation}
The energy per unit area in the right-hand pulse is
\begin{equation}
\frac{U_{r}}{\mathcal{A}}=\int_{z=ct-\pi/2k}^{z=ct+\pi/2k}dz\frac{1}{8\pi
}(\mathbf{E}_{r}^{2}+\mathbf{B}_{r}^{2})=\int_{z=ct-\pi/2k}^{z=ct+\pi/2k}%
\frac{dz}{8\pi}2E_{0}^{2}\cos^{2}(kz-\omega t)=\frac{1}{8k}E_{0}^{2},
\label{UrA}%
\end{equation}
and there is an equal energy per unit area in the left-hand pulse. \ 

The momentum per unit area $\mathbf{P}_{r}/\mathcal{A}$ carried in the
right-hand pulse is given by
\begin{equation}
\mathbf{P}_{r}/\mathcal{A}=\int_{z=ct-\pi/2k}^{z=ct+\pi/2k}dz\frac{1}{4\pi
c}\mathbf{E}_{r}\times\mathbf{B}_{r}=\widehat{z}\int_{z=ct-\pi/2k}%
^{z=ct+\pi/k}dz\frac{1}{4\pi c}E_{0}^{2}\cos^{2}(kz-\omega t)=\widehat{z}%
\frac{1}{8\omega}E_{0}^{2}. \label{PrA}%
\end{equation}
The momentum per unit area $\mathbf{P}_{l}/\mathcal{A}$ in the left-hand pulse
has equal magnitude but is opposite in direction so that the total
electromagnetic momentum vanishes.

The calculation which we have carried out corresponds to a traditional
calculation in classical electromagnetism. \ What we want to emphasize is that
the energy emitted by the surface current is symmetric on either side of the
emitting source and that there is no recoil of the radiation source. \ This is
the traditional view relevant when there is no interference with source-free radiation.

\section{Interference of a Source-Free Wave with a Wave Generated by a Current
Pulse}

\subsection{Source-Free Standing Plane Wave}

Now we come to the essential calculation of this article. \ In addition to the
radiation pulse generated by the pulsed current sheet, we will assume that
there is a source-free plane wave present in the space. \ Although the use of
a \textit{running} plane wave gives extremely interesting interference
behavior, we are particularly interested in particle-like behavior for waves.
\ Therefore we will assume the presence of an electromagnetic
\textit{standing} plane wave which has no preferred direction of propagation.
\ The source-free standing wave
\begin{equation}
\mathbf{E}_{S}(\mathbf{r,}t)=\widehat{x}E_{S}\cos(kz+\phi)\cos(\omega
t+\theta),\text{ \ \ \ }\mathbf{B}_{S}(\mathbf{r,}t)=\widehat{y}E_{S}%
\sin(kz+\phi)\sin(\omega t+\theta) \label{ESS}%
\end{equation}
exists throughout spacetime, where $\omega=ck,$ and the phases $\phi$ and
$\theta$ are fixed constants. \ 

We can easily check that the standing wave fields in Eq. (\ref{ESS}) indeed
satisfy the source-free Maxwell equations. \ The source-free standing
electromagnetic wave involves no time-average transfer of energy or momentum
in space. \ The energy density $u(\mathbf{r},t)$ does indeed oscillate in
time, since
\begin{align}
u_{S}(z,t)  &  =\frac{1}{8\pi}\left[  \mathbf{E}_{S}^{2}(\mathbf{r}%
,t)+\mathbf{B}_{S}(\mathbf{r}.t)\right] \nonumber\\
&  =\frac{E_{S}^{2}}{8\pi}\left[  \cos^{2}(kz+\phi)\cos^{2}(\omega
t+\theta)+\sin^{2}(kz+\phi)\sin^{2}(\omega t+\theta)\right]  . \label{uS}%
\end{align}
However, no energy crosses the nodes in the standing wave electric field,
since%
\begin{equation}
\mathbf{S}_{S}(z,t)=\frac{c}{4\pi}\mathbf{E}_{S}(\mathbf{r}.t)\times
\mathbf{B}_{S}(\mathbf{r,}t)=\widehat{z}\frac{cE_{S}^{2}}{16\pi}\sin
[2(kz+\phi)]\sin[2(\omega t+\theta)], \label{SS}%
\end{equation}
which vanishes at the electric field nodes at $kz+\phi=n\pi,$ for integer $n.$
\ The energy per unit area oscillates back and forth locally within half a
wavelength. \ 

\subsection{Superposition of the Fields of the Source-Free Standing Wave and
the Pulsed Current Sheet}

\subsubsection{Energy Aspects}

If the external agent causes a pulse in the surface current $\mathbf{K}(t)$ at
$z=0$ corresponding to Eq. (\ref{K}) during the time interval $-\pi
/(2\omega)<t<\pi/(2\omega)$ in the presence of the standing wave, then the
total electromagnetic fields in spacetime will correspond to the superposition
of the source-free fields in Eq. (\ref{ESS}) plus the fields generated by the
surface-current pulse as given in Eqs. (\ref{Erp}) and (\ref{Elp}).

The connection between the surface current amplitude $K_{0}$ and the amplitude
$E_{0}$ of the radiated fields remains that given in Eq. (\ref{EK}).
\ However, we notice that the power per unit area $\mathcal{P}/\mathcal{A}%
$\ delivered by the external agent now involves not only the pulse fields but
also the fields of the source-free standing wave. \ Thus we find%
\begin{align}
\mathcal{P}/\mathcal{A}  &  \mathcal{=}-\mathbf{K}(t)\cdot\left\{
\mathbf{E}_{S}(0,t)+\left[  \mathbf{E}_{r}(0,t)+\mathbf{E}_{l}(0,t)\right]
/2\right\} \nonumber\\
&  =-K_{0}\cos(\omega t)E_{S}\cos(\phi)\cos(\omega t+\theta)-K_{0}E_{0}%
\cos^{2}(\omega t)\nonumber\\
&  =-K_{0}E_{S}\cos^{2}(\omega t)\cos\phi\cos\theta+K_{0}E_{S}\cos(\omega
t)\sin(\omega t)\cos\phi\sin\theta-K_{0}E_{0}\cos^{2}(\omega t). \label{PASP}%
\end{align}
The work done by the external agent which produced the surface current
involves the time integral of the power per unit area in Eq. (\ref{PASP}), giving%

\begin{equation}
U/\mathcal{A=}\int_{t=-\pi/2\omega}^{t=\pi/2\omega}dt\frac{\mathcal{P}%
}{\mathcal{A}}=\frac{E_{S}E_{0}}{4k}\cos\phi\cos\theta+\frac{E_{0}^{2}}{4k},
\label{UASP}%
\end{equation}
where we have replaced $K_{0}$ by $(-c/2\pi)E_{0}$ from Eq. (\ref{EK}).

The electromagnetic fields in space involve the superposition $\mathbf{E}%
_{S}+\mathbf{E}_{r,l},~\mathbf{B}_{S}+\mathbf{B}_{r,l}$ of the standing wave
fields $\mathbf{E}_{S},~\mathbf{B}_{S}$ and the pulse fields $\mathbf{E}_{r},$
$\mathbf{B}_{r},~\mathbf{E}_{l},~\mathbf{B}_{l}$ emitted by the surface
current. \ The energy density in space corresponds to that of the standing
plane wave, except in the spatial region where the pulse fields are present.
\ The \textit{net} energy per unit area \textit{above} the energy per unit
area of the standing wave is given by%

\begin{align}
&  U_{R}(z,t)/\mathcal{A}\nonumber\\
&  \mathcal{=}\int_{z=ct-\pi/2k}^{z=ct+\pi/2k}dz\frac{1}{8\pi}\left\{  \left[
\mathbf{E}_{S}(z,t)+\mathbf{E}_{r}(z,t)\right]  ^{2}+\left[  \mathbf{B}%
_{S}(z,t)+\mathbf{B}_{r}(z,t)\right]  ^{2}-\left[  \mathbf{E}_{S}%
^{2}(z,t)+\mathbf{B}_{S}^{2}(z,t)\right]  \right\} \nonumber\\
&  =\int_{z=ct-\pi/2k}^{z=ct+\pi/2k}dz\frac{1}{8\pi}\left\{  2\mathbf{E}%
_{S}(z,t)\cdot\mathbf{E}_{r}(z,t)+2\mathbf{B}_{S}(z,t)\cdot\mathbf{B}%
_{r}(z,t)+\mathbf{E}_{r}^{2}(z,t)+\mathbf{B}_{r}^{2}(z,t)\right\}
\label{URSP}%
\end{align}
on the right-hand side of the current sheet. \ The integral involving the
interference term for the electric fields can be evaluated as%
\begin{align}
&  \int_{z=ct-\pi/2k}^{z=ct+\pi/2k}dz\frac{1}{8\pi}\left[  2\mathbf{E}%
_{S}(z,t)\cdot\mathbf{E}_{r}(z,t)\right] \nonumber\\
&  =\int_{z=ct-\pi/2k}^{z=ct+\pi/2k}dz\frac{1}{4\pi}\left[  E_{S}\cos
(kz+\phi)\cos(\omega t+\theta)E_{0}\cos(kz-\omega t)\right] \nonumber\\
&  =\frac{E_{S}E_{0}}{8k}\cos(\omega t+\theta)\cos(\omega t+\phi), \label{Int}%
\end{align}
and the integral involving the magnetic fields can be evaluated in a similar
fashion. \ We find that on the right-hand side of the current sheet, the
electromagnetic field energy per unit area above the energy per unit area of
the source-free standing wave is constant and is given by%
\begin{equation}
U_{R}/\mathcal{A=}\frac{E_{S}E_{0}}{8k}\cos(\phi-\theta)+\frac{E_{0}^{2}}{8k}.
\label{URASP}%
\end{equation}
Similarly, the electromagnetic field energy per unit area above the energy per
unit area of the source-free plane wave on the left-hand side of the current
sheet is
\begin{equation}
U_{L}/\mathcal{A=}\frac{E_{S}E_{0}}{8k}\cos(\phi+\theta)+\frac{E_{0}^{2}}{8k}.
\label{ULASP}%
\end{equation}

We notice that, as expected, the energy per unit area delivered by the
external agent given in Eq. (\ref{UASP}) fits exactly with the energy per unit
area moving out on either side of the current source,%
\begin{align}
U/\mathcal{A}  &  \mathcal{=}\frac{E_{S}E_{0}}{4k}\cos\phi\cos\theta
+\frac{E_{0}^{2}}{4k}\nonumber\\
&  \mathcal{=}\frac{E_{S}E_{0}}{8k}\cos(\phi-\theta)+\frac{E_{0}^{2}}%
{8k}+\frac{E_{S}E_{0}}{8k}\cos(\phi+\theta)+\frac{E_{0}^{2}}{8k}\nonumber\\
&  =U_{R}/\mathcal{A}+U_{L}/\mathcal{A}. \label{URL}%
\end{align}

\subsubsection{Momentum Aspects}

In addition to providing energy, the external agent must now provide an
external force per unit area acting on the surface current so as to balance
the electromagnetic force per unit area on the surface current. \ The net
force per unit area\ on the surface current arises from the standing wave
alone, since the magnetic field due to the current pulses averages to zero at
$z=0.~$The force per unit area due to the external agent must be given by
\begin{align}
\mathbf{F/}\mathcal{A}  &  \mathcal{=-}\frac{\mathbf{K(t)}}{c}\times\left\{
\mathbf{B}_{S}\mathbf{(}0,t)+\left[  \mathbf{B}_{r}(0,t)+\mathbf{B}%
_{l}(0,t)\right]  /2\right\}  =-\widehat{z}K_{0}\cos(\omega t)E_{S}\sin
(\phi)\sin(\omega t+\theta)\nonumber\\
&  =-\widehat{z}(K_{0}E_{S}/c)\left\{  \cos(\omega t)\sin(\omega t)\sin
(\phi)\cos(\theta)+\cos(\omega t)\cos(\omega t)\sin(\phi)\sin(\theta)\right\}
, \label{FASP}%
\end{align}
where we have used the magnetic field $\mathbf{B}_{S}(0,t)$ from Eq.
(\ref{ESS}). \ The net impulse $\mathbf{I}/\mathcal{A}$ per unit area
delivered by the external agent is given by the time integral of the external
force per unit area in Eq. (\ref{FASP}), giving from Eq. (\ref{EK})%
\begin{equation}
\mathbf{I}/\mathcal{A=}\int_{t=-\pi/2\omega}^{t=\pi/2\omega}dt\frac
{\mathbf{F}}{\mathcal{A}}=-\widehat{z}\frac{\pi}{2c\omega}K_{0}E_{S}\sin
\phi\sin\theta=\widehat{z}\frac{E_{S}E_{0}}{4\omega}\sin\phi\sin\theta.
\label{IASP}%
\end{equation}

We can compare this impulse with the linear momentum carried in the pulses
generated by the current sheet in the presence of the source-free standing
wave. \ The momentum per unit area carried in the region of the right-hand
electromagnetic pulse is given by%
\begin{align}
\mathbf{P}_{R}/\mathcal{A}  &  \mathcal{=}\int_{z=ct}^{z=ct+\pi/2k}dz\frac
{1}{4\pi c}\mathbf{E(}z,t)\times\mathbf{B}(z,t)\nonumber\\
&  \mathcal{=}\int_{z=ct-\pi/2k}^{z=ct+\pi/2k}dz\frac{1}{4\pi c}\left[
\mathbf{E}_{S}\mathbf{(}z,t)+\mathbf{E}_{r}(z,t)\right]  \times\left[
\mathbf{B}_{S}(z,t)+\mathbf{B}_{r}(z,t)\right] \nonumber\\
&  =\int_{z=ct-\pi/2k}^{z=ct+\pi/2k}dz\frac{1}{4\pi c}\left\{  \mathbf{E}%
_{S}\mathbf{(}z,t)\times\mathbf{B}_{S}(z,t)+\left[  \mathbf{E}_{r}%
(z,t)\times\mathbf{B}_{S}(z,t)+\mathbf{E}_{S}\mathbf{(}z,t)\times
\mathbf{B}_{r}(z,t)\right]  \right. \nonumber\\
&  \left.  +\mathbf{E}_{r}(z,t)\times\mathbf{B}_{r}(z,t)\right\}  \label{PR}%
\end{align}
Once again, it is easy to evaluate the integrals. \ We find
\begin{align}
&  \int_{z=ct-\pi/2k}^{z=ct+\pi/2k}dz\frac{1}{4\pi c}\mathbf{E}_{S}%
\mathbf{(}z,t)\times\mathbf{B}_{S}(z,t)\nonumber\\
&  =\widehat{z}\int_{z=ct-\pi/2k}^{z=ct+\pi/k}dz\frac{1}{4\pi c}\cos
(kz+\phi)\cos(\omega t+\theta)\sin(kz+\phi)\sin(\omega t+\theta)=0, \label{I2}%
\end{align}
so that the standing wave itself carries no net linear momentum when summed
over half a wavelength. \ Next we have%
\begin{align}
&  \int_{z=ct-\pi/2k}^{z=ct+\pi/2k}dz\frac{1}{4\pi c}\left[  \mathbf{E}%
_{r}(z,t)\times\mathbf{B}_{S}(z,t)+\mathbf{E}_{S}\mathbf{(}z,t)\times
\mathbf{B}_{r}(z,t)\right] \nonumber\\
&  =\widehat{z}\int_{z=ct-\pi/2k}^{z=ct+\pi/2k}dz\frac{E_{S}E_{0}}{4\pi
c}\left[  \cos(kz-\omega t)\sin(kz+\phi)\sin(\omega t+\theta)\right.
\nonumber\\
&  \left.  +\cos(kz+\phi)\cos(\omega t+\theta)\cos(kz-\omega t)\right]
\nonumber\\
&  =\widehat{z}\frac{E_{S}E_{0}}{8\omega}\cos(\phi-\theta). \label{I3}%
\end{align}
The integral involving $\mathbf{E}_{r}$ and $\mathbf{B}_{r}$ was given in Eq.
(\ref{PrA}). \ Thus we have finally
\begin{equation}
\mathbf{P}_{R}/\mathcal{A=}\widehat{z}\frac{E_{S}E_{0}}{8\omega}\cos
(\phi-\theta)+\widehat{z}\frac{1}{8\omega}E_{0}^{2}, \label{PRSP}%
\end{equation}
so that the momentum on the right-hand side of the current sheet is constant
in time and involves an interference between the source-free plane wave and
the pulse created by the current sheet. \ An analogous calculation involving
the pulse on the left gives%
\begin{equation}
\mathbf{P}_{L}/\mathcal{A=-}\widehat{z}\frac{E_{S}E_{0}}{8\omega}\cos
(\phi+\theta)-\widehat{z}\frac{1}{8\omega}E_{0}^{2}. \label{PLSP}%
\end{equation}
As expected, there is momentum balance; the net impulse per unit area
delivered by the external agent in the presence of the standing plane wave was
given in Eq. (\ref{IASP}), so that
\begin{align}
\mathbf{I}/\mathcal{A}  &  =\widehat{z}\frac{E_{S}E_{0}}{4\omega}\sin\phi
\sin\theta\nonumber\\
&  =\widehat{z}\frac{E_{S}E_{0}}{8\omega}\cos(\phi-\theta)+\widehat{z}\frac
{1}{8\omega}E_{0}^{2}-\widehat{z}\frac{E_{S}E_{0}}{8\omega}\cos(\phi
+\theta)-\widehat{z}\frac{1}{8\omega}E_{0}^{2}\nonumber\\
&  =\mathbf{P}_{R}/\mathcal{A+}\mathbf{P}_{L}/\mathcal{A}. \label{IRL}%
\end{align}

\subsection{Examples of Non-uniform Spreading of the Radiation Pulse Energy in
the Presence of Source-Free Radiation}

\subsubsection{Symmetrical Outgoing Pulses of Positive Energy for Large
Radiation Pulses, $E_{0}>>E_{S}$}

In the absence of the standing plane wave, the current sheet generates pulses
(\ref{Erp}) and (\ref{Elp}) which carry energy and momentum away from the
sheet in a symmetrical fashion, as given in equations (\ref{UrA}) and
(\ref{PrA}). \ However, in the presence of the standing plane wave, the energy
and momentum carried in the pulses depends upon the phases in both space and
time of the standing wave relative to the oscillation of the current sheet,
and also on the relative magnitude of the standing wave field and of the field
generated by the surface current. \ If the magnitude $E_{0}$ of the field
generated by the surface current is very large compared to the field $E_{S}$
of the standing wave, then the standing wave can be ignored; the pulses
generated by the surface current carry equal amounts of energy and momentum,
spreading out symmetrically from the current source. \ There is no recoil of
the sources of radiation. \ This corresponds to the same situation where the
source-free plane wave is absent, and is the situation generally assumed in
the physics literature.

\subsubsection{Small Radiation Pulses, $E_{S}>>E_{0}$}

If the magnitude $E_{S}$ of the plane wave is much larger than the field
$E_{0}$ generated by the current source, then the interference terms involving
$E_{S}E_{0}$ dominate the situation, while the terms in $E_{0}^{2}$ are
negligible. \ The interference terms in $E_{S}E_{0}$ depend upon the relative
phases $\phi$ and $\theta.$

\paragraph{Case $\phi=0,~\theta=0,~E_{S}>>E_{0}$}

If $\phi=0$ while $\theta=0,$ then the current source introduces pulses of
large positive energy which move symmetrically outwards from the current
sheet. \ The external agent introduces an energy per unit area in Eq.
(\ref{UASP}) $U/\mathcal{A\approx\lbrack}E_{S}E_{0}/(4k)],$ but no net
momentum in Eq. (\ref{IASP}). \ If the spatial phase is $\phi=0$ while the
temporal phase is taken as $\theta=\pi,$ then the current source generates
large pulses of negative relative energy per unit area (holes in the standing
wave energy) which travel outwards symmetrically. \ The energy per unit area
delivered by the external agent from Eq. (\ref{UASP}) is $U/\mathcal{A\approx
-[}E_{S}E_{0}/(4k)],$ but again no net momentum is delivered. \ The pulses of
radiation carry negative relative energy (holes in the standing wave) from
Eqs. (\ref{URASP}) and (\ref{ULASP}) $U_{R}=U_{L}\approx-E_{S}E_{0}/(8k)$ and
momentum in Eq. (\ref{PRSP}) $\mathbf{P}_{R}/\mathcal{A\approx-}%
\widehat{z}E_{S}E_{0}/(8\omega)$ and in Eq. (\ref{PLSP}) $\mathbf{P}%
_{L}/\mathcal{A\approx}\widehat{z}E_{S}E_{0}/(8\omega);$ the negative energy
moves outwards from the current sheet, but the momentum travels in the
opposite direction from the direction of negative energy motion. \ The energy
behaves as a \textquotedblleft hole\textquotedblright\ in the standing wave
energy. \ 

\paragraph{Case $\phi=\pi/2,$ $\theta=\pi/2,$ $E_{S}>>E_{0}$}

\ If the spatial and temporal phases of the plane wave are given by $\phi
=\pi/2$ and $\theta=\pi/2,$ then the current pulse has introduced no net
electromagnetic energy per unit area into the system since $U/\mathcal{A}%
\approx0$ in Eq. (\ref{UASP}), but the current pulse has introduced net linear
momentum since $\mathbf{I}/\mathcal{A}=\widehat{z}E_{S}E_{0}/4\omega$ in Eq.
(\ref{IASP}). \ In this case, a positive energy per unit area travels to the
right as in Eq. (\ref{URASP}) with $U_{R}/\mathcal{A\approx}E_{S}E_{0}/(8k)$
carrying linear momentum as in Eq. (\ref{PRSP}) $\mathbf{P}_{R}%
/\mathcal{A\approx}\widehat{z}E_{S}E_{0}/(8\omega)$, while a negative relative
energy per unit area as in Eq. (\ref{ULASP}) with $U_{L}/\mathcal{A\approx
-}E_{S}E_{0}/(8k)$ travels to the left acting like a \textquotedblleft
hole\textquotedblright\ in the plane wave carrying negative relative energy to
the left and momentum\ from Eq. (\ref{PLSP}) as $\mathbf{P}_{L}%
/\mathcal{A\approx}\widehat{z}E_{S}E_{0}/(8\omega)$\ to the right. \ If the
sign of the temporal phase angle is reversed so $\theta=-\pi/2,$ then the
directions taken by the positive and negative energy pulses are reversed.

\subsubsection{Radiation Pulses Matching the Source-Free Radiation:
Particle-Like Behavior}

\paragraph{Case $\phi=\pi/2,$ $\theta=\pi/2,$ $E_{S}=E_{0}$}

Perhaps the most interesting situation involves the case when the field
$E_{0}$ of the pulse is of the same order as the amplitude $E_{S}$ of the
standing wave, so that $E_{0}\approx E_{S}.$ \ In this case, if we have
$\phi=\pi/2$ and $\theta=\pi/2,$ while $E_{S}=E_{0},$ then we can have
\textit{one} pulse of positive energy from Eq. (\ref{UASP}) with
$U/\mathcal{A=}$ $E_{0}^{2}/(4k)$ which is moving to the right, according to
Eq. (\ref{URASP}) with momentum given by Eq. (\ref{PRSP}) $\mathbf{P}%
/\mathcal{A=}$ $\widehat{z}E_{0}^{2}/(4\omega).$ \ There is no pulse to the
left since Eqs. (\ref{ULASP}) and (\ref{PLSP}) both give vanishing
contributions. \ Under this situation, the current sheet current recoils to
the left and can be held in place by the impulse per unit area to the right
exerted by the external agent, as in Eq. (\ref{IASP}) with $\mathbf{I}%
/\mathcal{A=}\widehat{z}E_{0}^{2}/(4\omega)$. \ If the phases are $\phi=\pi/2$
and $\theta=-\pi/2,$ then, the equations indicate that there is one
electromagnetic pulse moving to the left while the current sheet recoils to
the right. \ Qualitatively, this is the sort of behavior which is interpreted
in terms of photons for the electromagnetic pulses. \ Energy and momentum are
conserved, but the asymmetry of the radiation emission in association with the
corresponding source recoil depends upon the existence of source-free
radiation which is present at the time of the radiation emission. \ 

\subsubsection{Average Behavior for Random Source-Free Radiation}

In the discussions above, we have considered fixed phases in space and time
for the standing wave. $\ $However, for thermal radiation or zero-point
radiation, the appropriate description involves random phases. \ If we regard
the phases $\phi$ and $\theta$ as independent random phases and average over
the random phases, then the interference terms in Eqs. (\ref{UASP}),
(\ref{URASP}), (\ref{ULASP}), (\ref{IASP}), (\ref{PRSP}), and (\ref{PLSP}) all
vanish. \ Thus when averaged over random phases, the average effect
corresponds to electromagnetic waves spreading symmetrically out from the
current source and gives no evidence of the particle-like behavior which holds
for specific choices of relative phase. \ 

\section{Discussion}

Although the current textbooks of classical electromagnetism do not mention
the source-free radiation terms appearing in Eqs. (\ref{V}) and (\ref{A}), the
accurate classical description of nature must include the source-free terms.
\ The source-free radiation is crucial for understanding the behavior of a
broadcasting antenna located in the radio waves from another source. \ Also,
the source-free random radiation corresponding to classical zero-point
radiation provides classical insights into aspects of modern physics. It has
been pointed out that the presence of source-free classical electromagnetic
zero-point radiation (as random classical radiation with a Lorentz invariant
spectrum) allows us to understand a number of phenomena which are usually
claimed to have explanations only in terms of energy quanta.\cite{any} \ Thus
if classical zero-point radiation is included as source-free random radiation,
then there are natural classical explanations for Casimir forces, van der
Waals forces, diamagnetism, blackbody radiation, and the absence of
\textquotedblleft atomic collapse.\textquotedblright\ \ 

It was shown over 45 years ago that the \textit{fluctuations} of thermal
radiation could be understood purely classically, without photons, in terms of
classical waves when blackbody radiation included classical zero-point
radiation.\cite{fluct} \ The present simple model calculation raises the
possibility that what is currently regarded as individual photon behavior may
perhaps also be understood in terms of conventional classical electrodynamics.
\ Thus if zero-point radiation is assumed to be present, then any current
source will interact with this source-free radiation, just as illustrated in
our simple example. \ Since zero-point radiation is \textit{random} radiation,
the total \textit{angular} radiation pattern corresponds to that found by
averaging over all the random phases, and will behave symmetrically, as though
the zero-point radiation interference were not present. \ However, the
individual pulses of radiation will indeed be altered by the presence of
zero-point radiation. \ If the current source generates an electromagnetic
pulse where the electric field is much larger than the radiation fields of the
random zero-point radiation in the same frequency region, then the presence of
zero-point radiation can be ignored. \ This is the situation for macroscopic
radiating antennas. \ However, if the current source generates an
electromagnetic pulse where the electric field is comparable to the radiation
fields of the zero-point radiation in the same frequency region, then the
interference between the zero-point radiation and the radiation generated by
the current source will be important. \ The idea of photons was taken up
convincingly in connection with x-ray behavior which showed both classical
interference effects in connection with Bragg scattering but also photon-like
behavior in connection with Compton scattering. \ Of course, x-rays involve
high frequency radiation where the zero-point radiation per normal mode
$U_{0}(\omega)=(1/2)\hbar\omega$ is large and the wave-trains of emitted
radiation are small. \ These are exactly the conditions where our simple
example suggests one would find convincing particle-like behavior for the
classical radiation pulses.

In this article, we have given a very simple example for the interference of
source-free electromagnetic radiation with the radiation generated by a
current source. \ The physics and mathematics of the example are at the level
of an undergraduate text. \ However, the calculation may suggest a way of
understanding the particle-like behavior of radiation pulses purely within the
conventional wave theory of classical electrodynamics.

\section{Acknowledgement}

I wish to thank Professor David J. Griffiths for helpful suggestions regarding
this work.


\begin{thebibliography}{9}                                                                                                %


\bibitem {gaussian}We are using gaussian units which are natural to electrodynamics.

\bibitem {Griffiths}D. J. Griffiths, \textit{Introduction to Electrodynamics}
4th edn (Pearson, New York 2013), p. 445.

\bibitem {Jackson}J. D. Jackson, \textit{Classical Electrodynamics} (John
Wiley \& Sons, New York, 1999), 3rd ed., p. 246. \ 

\bibitem {Zangwill}A. Zangwill, \textit{Modern Electrodynamics} (Cambridge U.
Press, 2013), p. 725.

\bibitem {Garg}A. Garg, \textit{Classical Electromagnetism in a Nutshell}
(Princeton U. Press, Princeton, NJ 08450, 2012), p. 204.

\bibitem {G499}See, for example, Griffiths in ref. 2, p. 499, Problem 11.28.

\bibitem {G397}See, for example, Griffiths in ref. 2, p. 397.

\bibitem {any}T. H. Boyer, \textquotedblleft Any classical description of
nature requires classical electromagnetic zero-point
radiation,\textquotedblright\ Am. J. Phys. \textbf{79}, 1163-1167 (2011);
"Understanding zero-point energy in the context of classical
electromagnetism," Eur. J. Phys. \textbf{37}, 055206(14) (2016). \ 

\bibitem {fluct}T. H. Boyer, \textquotedblleft Classical Statistical
Thermodynamics and Electromagnetic Zero-Point Radiation,\textquotedblright%
\ Phys. Rev. \textbf{186}, 1304-1318 (1969).
\end{thebibliography}
\end{document}